\documentclass[preprint,aps]{revtex4-1}
\usepackage{graphicx}
\usepackage{subfig}
\usepackage{caption}
\usepackage{epstopdf}
\usepackage{amsfonts}
\usepackage{amsmath}
\usepackage{amssymb}
\begin{document}


\title{Tunable Electron Multibunch Production in Plasma Wakefield Accelerators}


\author{B. Hidding$^{1,2,3}$, O. Karger$^2$, G. Wittig$^2$, C. Aniculaesei$^2$, D. Jaroszynski$^1$, B.W.J. McNeil$^1$, L.T. Campbell$^2$, M.R. Islam$^1$, B. Ersfeld$^1$, Z.-M. Sheng$^1$, Y. Xi,$^3$, A. Deng$^3$, J.B. Rosenzweig$^3$, G. Andonian$^{3,4}$, A. Murokh$^4$, M.J. Hogan$^5$, D.L. Bruhwiler$^{6,7}$, E. Cormier$^8$}

\affiliation{$^1$Department of Physics, SUPA, Strathclyde University, Glasgow, UK, G4 0NG, $^2$Department of Experimental Physics, University of Hamburg \& CFEL, $^3$  Department of Physics and Astronomy, University of California, Los Angeles, USA, $^4$  RadiaBeam Technologies, Santa Monica, USA, $^5$  SLAC, Stanford, USA, $^6$ University of Colorado at Boulder, Center for Integrated Plasma Studies, Boulder, Colorado 80309, USA, $^7$ RadiaSoft LLC, Boulder, Colorado 80304, USA, $^8$ Tech-X Corp., 5621 Arapahoe Ave., Boulder 80303, USA}

\date{\today}

\begin{abstract} 
Synchronized, independently tunable and focused $\mu$J-class laser pulses are used to release multiple electron populations via photo-ionization inside an electron-beam driven plasma wave. By varying the laser foci in the laboratory frame and the position of the underdense photocathodes in the co-moving frame, the delays between the produced bunches and their energies are adjusted. The resulting multibunches have ultra-high quality and brightness, allowing for hitherto impossible bunch configurations such as spatially overlapping bunch populations with strictly separated energies, which opens up a new regime for light sources such as free-electron-lasers. 
\end{abstract}

\maketitle
Plasma wakefield acceleration \cite{Chen1985PhysRevLett.54.693,PhysRevLett.61.98Rosenzweig1988} offers the advantage of ultrahigh accelerating gradients of the order of tens of GV/m or more, which allows for acceleration of particles to ultrarelativistic energies within few cm -- in contrast with conventional accelerators based on metallic cavities, where breakdown limits the accelerating fields to tens of MV/m. Another fundamental difference is that in conventional accelerators a large number of cavities 
are needed, in contrast to plasma accelerators where one transient plasma-cavity in the co-moving frame is sufficient. This plasma cavity is generated on the fly, either by an intense electron beam or a laser pulse with a duration $\tau < \lambda_p/2$, where $\lambda_p$ is the relativistic plasma wavelength, which is equal to the plasma cavity length.  
GeV-scale energies have been demonstrated both in laser wakefield accelerators (LWFAs) \cite{LeemansNatPhys2006,KarschNJP,WangTexas2GeVNatComm2013} and in beam-driven plasma wakefield accelerators (PWFAs) in cm to meter-scale \cite{Blumenfeld2007} plasma sections.
Witness bunch(es) to be accelerated in these systems can be injected from an external source or generated directly within the plasma itself. In this Letter, we show how ultra-high quality electron bunch trains can be simultaneously generated in one and the same plasma cavity  and tuned independently over a large parameter range. Such highly tunable multibunches have been hitherto inaccessible but are highly important for various fields, such as driver-witness plasma wakefield acceleration, and light sources based on Compton scattering or (multi-color \cite{PhysRevLett.72.2387Two-ColorFELJaroszynski1994}) free-electron lasers (FELs). 

\begin{figure}[b]
\begin{center}
 \includegraphics[width=0.79\textwidth]{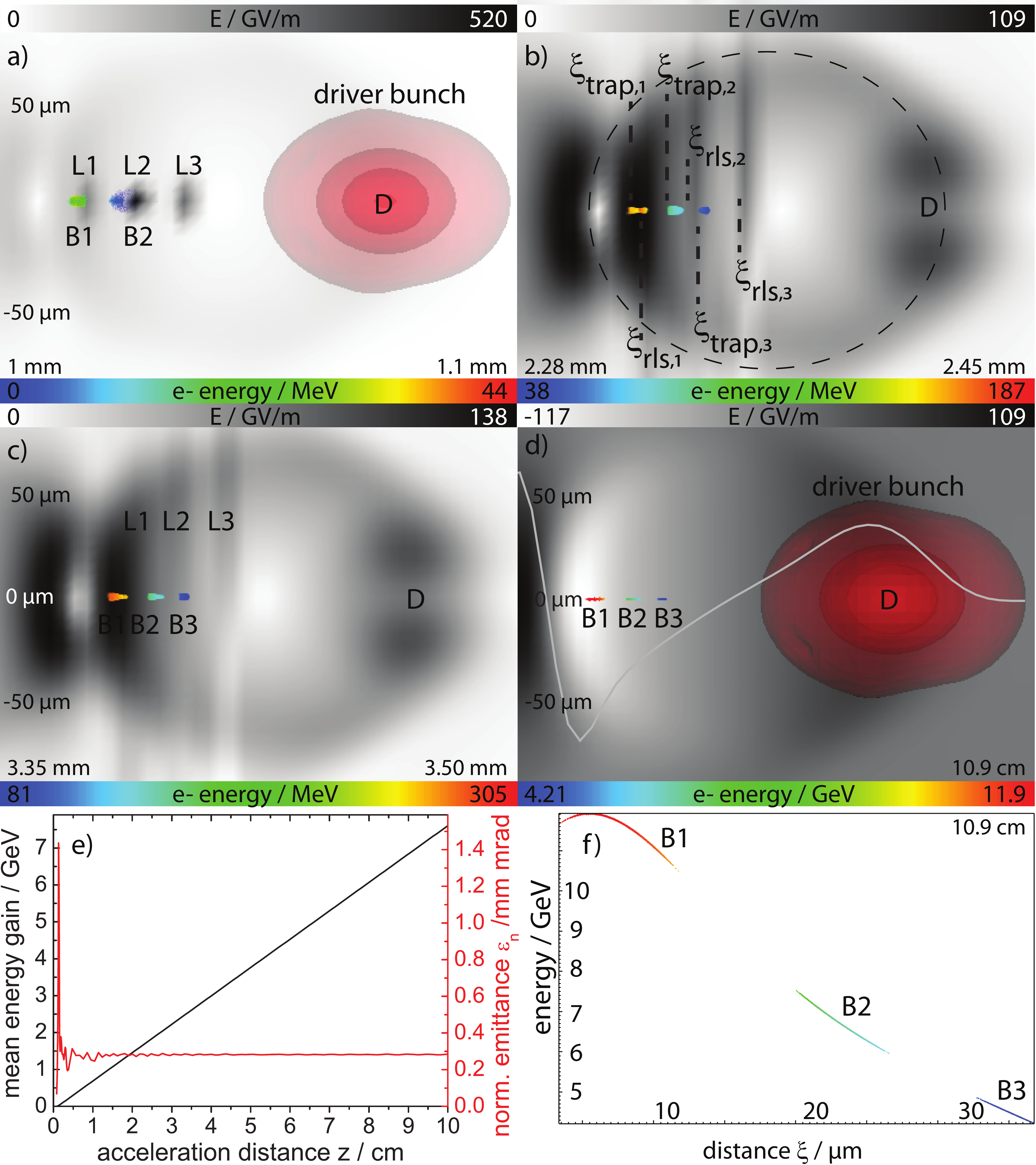} 
\end{center}
\caption{3D-PIC simulations of triple-bunch B1--B3 generation with three laser pulses L1--L3. 
In a)-c) the sum of the electric field shows the $y$-polarized L1--L3 whereas in d) the longitudinal component of the quasi-static system is shown, characterized by linear energy gain and asymptotic emittance (e), and strictly correlated longitudinal phase space (f).}\label{trojanmultirls}
\end{figure}

Recently, a novel underdense photocathode scheme has been introduced which decouples the witness bunch generation from  plasma wave generation \cite{hiddingpatent2011,HiddingPRL2012PhysRevLett.108.035001,YunfengPhysRevSTAB.16.031303}. In this ``Trojan Horse'' scheme \cite{hidding:570beyond}, the electron beam driver sets up a plasma wave based on a low-ionization-threshold (LIT) plasma component. Then, a laser pulse, either in collinear geometry \cite{HiddingPRL2012PhysRevLett.108.035001,hiddingpatent2011}, at an arbitrary angle \cite{hiddingpatent2011}, or in colliding pulse perpendicular geometry \cite{LiPhysRevLett.111.015003PRL2013}) is focused into the plasma wave to ionize a higher-ionization-threshold (HIT) component that has not been previously ionized by the driver bunch or by preionization. Ultracold electrons are then released  in a tunable region around the laser focus position $z_{\mathrm{rls}}$ in the lab frame via tunnelling and/or multiphoton ionization \cite{YunfengPhysRevSTAB.16.031303}.  
Depending on the release position $\xi_{\mathrm{rls}} = z_{\mathrm{rls}},-v_{ph}t$ in the co-moving frame of the beam-driven plasma blowout, which propagates with a phase velocity $v_{ph}\approx c$, the plasma wake potential  $\phi$ may trap electrons at positions $\xi_{\mathrm{trap}}$. 
If the blowout is especially strong due to a large driver beam current $I_d$, the trapping region is large and electrons released at different $\xi_{\mathrm{rls}}$ will end up at different trapping positions $\xi_{\mathrm{trap}}$. Using the 3d particle-in-cell code VSIM \cite{Nieter2004448}, we explore the strategy to deploy multiple laser pulses within the first wave bucket in order to produce distinct accelerated electron populations that may differ in energy and/or space.  To this end, we use an envelope approximation to describe the release laser pulse(s), avoiding computationally expensive resolution of the laser pulse oscillations, and a static current to model the drive beam $I_d$. This approach does not capture the full physics of transverse oscillations imparted by the laser pulse, but is ideally suited to explore the bunch generation and acceleration over multi-cm scale distances to 10+ GeV energies. Furthermore, it offers additional insights into betatron phase mixing effects \cite{YunfengPhysRevSTAB.16.031303,PhysRevLett.112.035003Xu2014} independent of the residual momentum of the laser oscillation kick.   
For further simulation speed-up, we use Li as the LIT species and Li$^{+}$ as the HIT component to examine the underlying scheme. Using different LIT/HIT media such as H and He \cite{HiddingPRL2012PhysRevLett.108.035001} is possible without loss of generality \cite{hiddingpatent2011}. 
 The requirements of the laser pulses for releasing electrons locally confined within the blowout region are straightforward: a) they should be strongly focused and have a Rayleigh range $Z_R = \pi w_0^2 / \lambda < \lambda_p$, where $w_0$ is the beam waist and $\lambda$ the laser wavelength, and b) the focused laser intensities must be sufficient to ionize the HIT component. An upper limit is given by the barrier suppression ionization (BSI) threshold \cite{AugstJOSAB1991}, 
 which in case of Li$^{+}$ with an ionization energy of $W_{\mathrm{ion}} \approx 75.6$ eV corresponds to $I_{\mathrm{BSI}} \approx 4 \times 10^{9} \times W_{\mathrm{ion}}^4/(Z^2) \approx  3.2 \times10^{16}\,\mathrm{W/cm^{-2}}$ and an  $a_0 \approx \sqrt{\frac{I\lambda^2[\mathrm{\mu m^2}]}{1.37\times10^{18}\,\mathrm{V/m} }} \approx 0.12$, where $Z$ is the ionic charge and $\lambda = 0.8\, \mathrm{\mu m}$.  
As a basic principle, low $W_{\mathrm{ion}}$ and short $\lambda$ are desirable \cite{hidding:570beyond} to reduce the peak electric fields required for ionization. This is important because the electric field of the HIT ionizing laser pulse determines the residual momentum of the newly born electrons, because pre-ionization or self-ionization of the LIT medium is easier, and because the gap to the HIT level may be larger.  

Figure \ref{trojanmultirls} shows the release, trapping and acceleration of three electron bunches (B1-B3) generated by three consecutive laser pulses (L1-L3) in a strong blowout region set up by a FACET-class drive beam D with a size $\sigma_z = \sigma_r = 20\,\mathrm{\mu m}$ and charge of $Q \approx 3$ nC in a Li plasma. The LIT plasma density $n_e \approx 1.5 \times 10^{17}\,\mathrm{cm^{-3}}$ 
 corresponds to a plasma wavelength $\lambda_p \approx 75\,\mathrm{\mu m}$. Each laser pulse has a duration of $\tau = 25$ fs (FWHM), and they follow D with intervals of $\Delta \xi = 20\,\mathrm{\mu m}$. In the lab frame, the spots with $w_0 = 5\,\mathrm{\mu m}$ FWHM at an intensity of $a_0 = 0.14$ are separated by $\Delta z = 20\,\mathrm{\mu m}$; these are the release positions of the HIT electron populations.

In figure \ref{trojanmultirls} a), the peak electric field of $E_{\mathrm{sum}} \approx$ 520 GV/m is produced by the focused L2, which is in the process of liberating HIT electrons which form bunch B2, while L3 has not yet reached its focus. 
The trapping positions can be predicted from the potential differences $\psi_{\mathrm{trap}} - \psi_{\mathrm{rls}} = -1 + \sqrt{1+P_\perp^2}/\gamma_\mathrm{d}$ \cite{PakPRL2010PhysRevLett.104.025003,ChenIonizationTrappingPoP2012}, where $\psi = \phi - (v_{ph}/c)A_z$, $P_\perp = a_{\mathrm{\perp,rls}}$ is the canonical momentum, $A$ is the vector potential and $\gamma_\mathrm{d}$ is the drive beam Lorentz factor, and time is normalized to $\omega_p^{-1}$, length to $c/\omega_p$, momentum to $mc$ and the potential to $mc^2/|e|$. The sphericity of the blowout region is higher if the wake potential is large, and in the present case for an approximately spherically symmetric blowout region the potential can be modelled as $\psi = (r^2-R^2)/4$ where $r = \sqrt{\xi^2 + y^2+x^2}$ \cite{PhysRevLett.103.175003KostyukovTrappingPRL2009}. For electrons born on axis at $\xi_{\mathrm{rls}}$, for $P_\perp \le mc^2$ and $\gamma_d \gg 1$ the trapping position is $\xi_{\mathrm{trap}} \approx -\sqrt{4+\xi^2_{\mathrm{rls}}}$, where $\xi = 0$ is the blowout region center. 
This is reflected in figure \ref{trojanmultirls} b), where the underdense photocathode action is complete and bunches B1--B3 are formed and trapped. While the laser pulses continue to diffract, the accelerating system is quasistatic and electrons are accelerated (fig. \ref{trojanmultirls} b-d) with linear energy gain. 
The evolution of the energy gain $\Delta E$  and the normalized emittance $\epsilon_n$, averaged over all three bunches, is shown in fig. \ref{trojanmultirls} e). Although L1--L3 are approximated by envelopes, the saturated emittance of $\epsilon_n \approx  0.3\, \pi$ mm mrad observed in the simulations is in good agreement with the crude scaling for the laser pulse contribution to the emittance in \cite{HiddingPRL2012PhysRevLett.108.035001}, which would predict $\epsilon_n \approx w_0 a_0/2^{3/2} \approx 0.2\, \pi$ mm mrad. The emittance would be further decreased using HIT media with lower ionization thresholds, such as He. In fig. \ref{trojanmultirls} f) we show the corresponding longitudinal phase space distribution of the triple bunch after $z=10.9$ cm in this quasistatic system (see supplemental material movie), where B2 and B3 have a correlated energy chirp $\partial W/\partial \xi < 0$, but for B1, located at the end of the blowout region, in the first part $\partial W/\partial \xi > 0$, while in the second part $\partial W/\partial \xi < 0$. The peaked energy structure of B1 reflects the fact that this bunch resides in the accelerating electric field mimimum $E$, which opens up strategies for energy spread minimization.                   

\begin{figure}[htpb]
\begin{center}
  \includegraphics[width=0.99\textwidth]{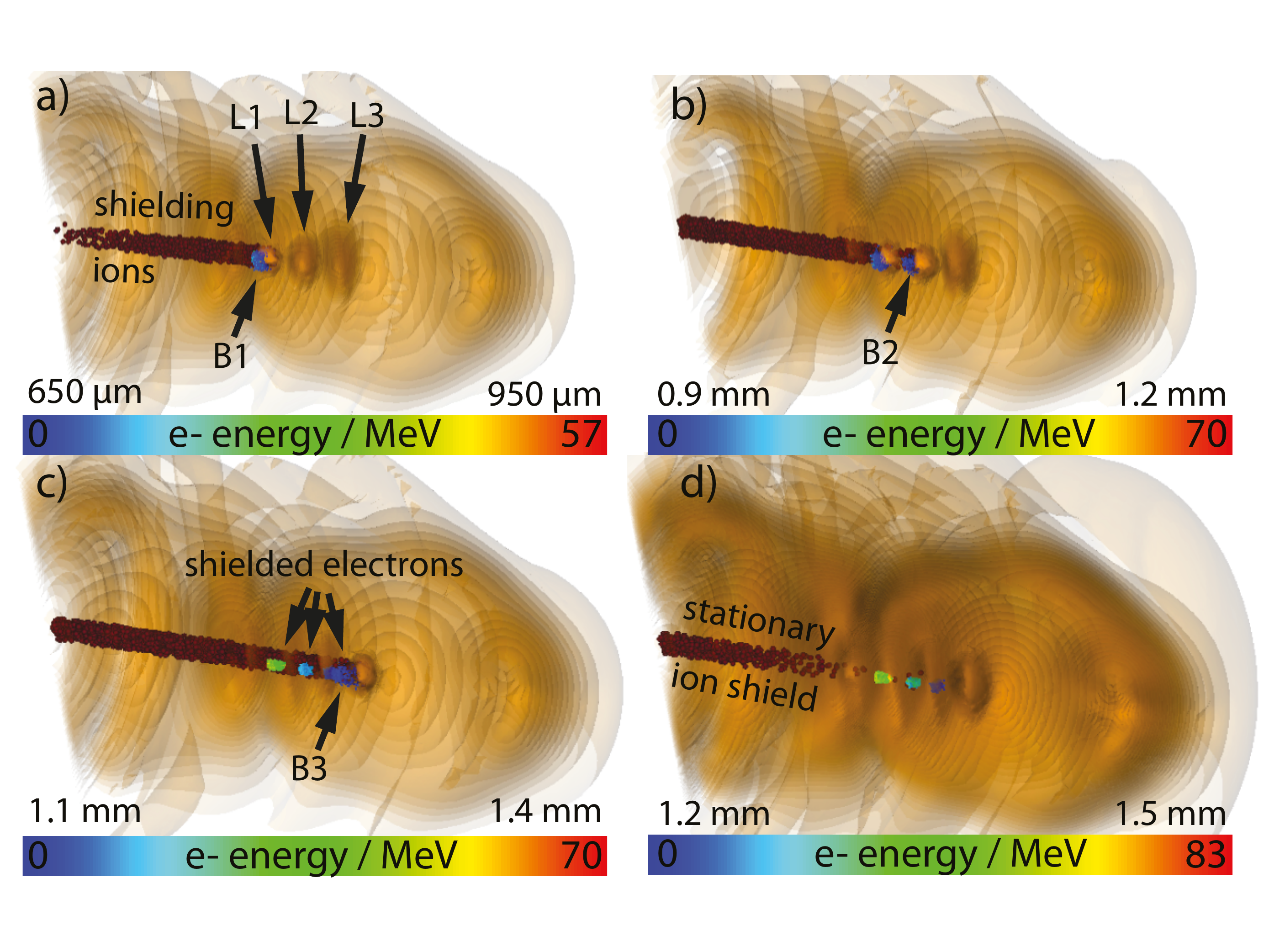}
\end{center}
\caption{Transient ion shielding. HIT ions (brown spheres) shield the forming electron bunches B1--B3 produced consecutively by laser pulses L1--L3 with $z_{\mathrm{rls,1}} < z_{\mathrm{rls,2}} < z_{\mathrm{rls,3}}$ to increase the shielding effect. }\label{transientshielding}
\end{figure}

It is advisable to release the electron bunches sequentially so that the electron bunch with the lowest release position $\xi_{\mathrm{rls}}$ is produced first and those with higher $\xi_{\mathrm{rls}}$ come later such that $\xi_{\mathrm{rls,n}} < \xi_{\mathrm{rls,n+1}}$, where $n$ denotes the laser pulse number. 
Also, if compatible with the bunch energy and delay parameters aimed at, the release of the electrons should happen in rapid succession, i.e. with minimized distances between the focus positions $\Delta z_n$. This way, the HIT electrons can profit the most from the transient space charge shielding effect of the HIT ions, which is an inherent feature of the underdense photocathode mechanism \cite{hidding:570beyond}. Furthermore, they do so at relatively low electron energies $\gamma$ where space charge forces would have the strongest detrimental effect. The ion shielding effect is visualized in figure \ref{transientshielding} by the contour plots of the electric field sum, sliced open to allow insight into the field distribution inside the blowout. In figure \ref{transientshielding} a), L1 has just reached the HIT and liberates electrons as well as Li$^{++}$ ions, which are shown with brownish spheres. These ions are quickly left behind, but provide a transient space charge shielding to the forming B1. In this snapshot, L2 and L3 are still in the focusing phase. In figure \ref{transientshielding} b), the second laser pulse has reached focus and releases HIT electrons and ions, which provide space charge shielding for the forming B2 and then also for the already formed B1, which still has a low energy. In figure \ref{transientshielding} c), the same happens with L3 and B3, and in figure \ref{transientshielding} d) the complete ionization and shielding process is complete. Each ion shield has a longitudinal  shape with a volume of approximately $V_{\mathrm{shield}} \approx Z_R \pi {(w_0/2)}^2$.  

\begin{figure}[htpb]
\begin{center}
  \includegraphics[width=0.99\textwidth]{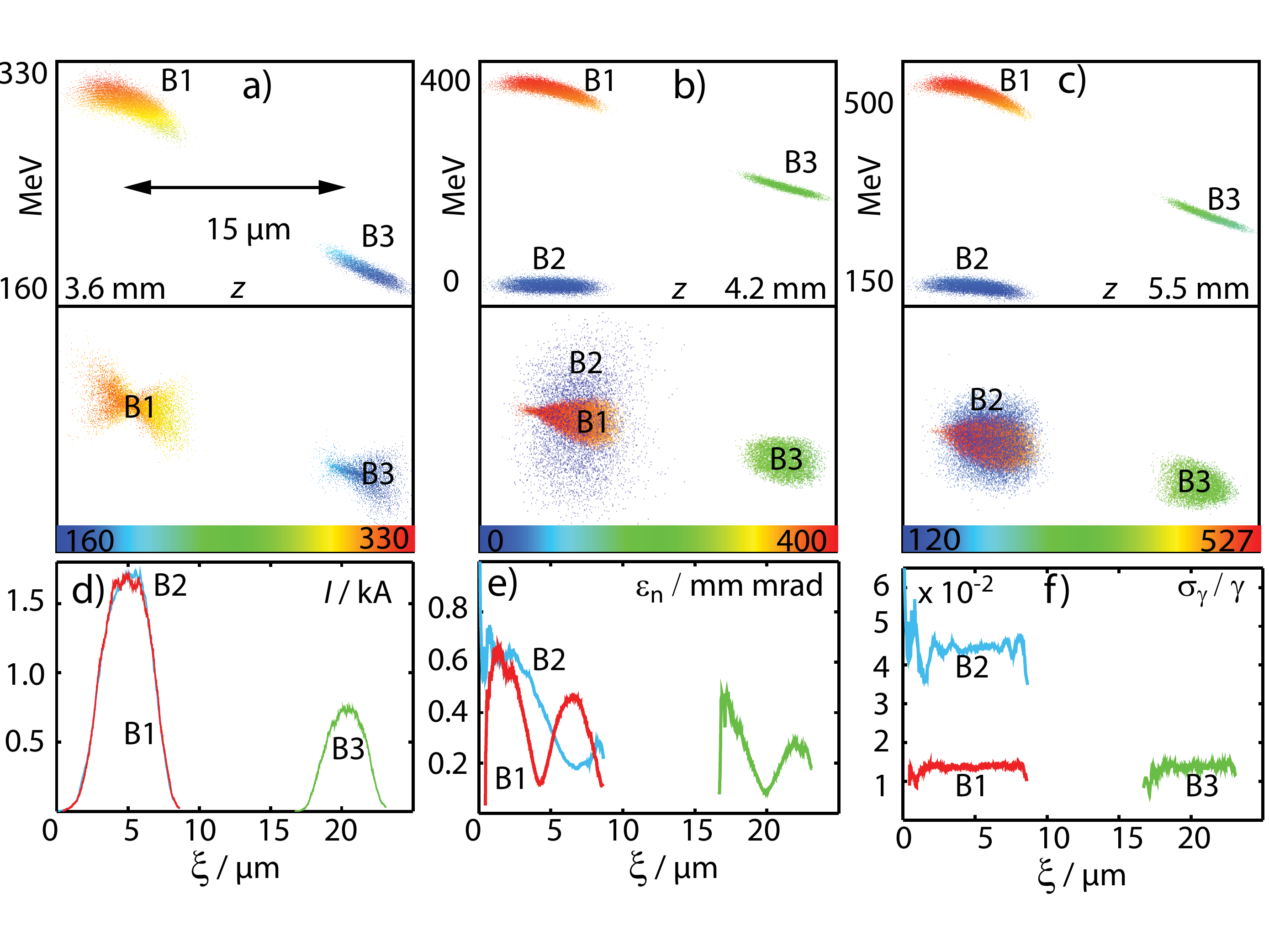}
\end{center}
\caption{Longitudinal phase space of three electron populations B1--B3 where B1 and B2 overlap in space but have distinct energies (a-c). For each population, the slice $I$ (d), $\epsilon_n$ (e) and $\sigma_\gamma / \gamma$ (f) after 5.5 mm acceleration are given.}\label{hotcoldpopulation}
\end{figure}

This multi-bunch generation scheme allows for electron bunch shaping and to produce highly exotic electron bunch configurations. For example, consider the case when two laser pulses have the same $\xi_{\mathrm{rls,n}}$ in the co-moving frame but different $z_{\mathrm{rls,n}}$ in the lab frame. The result is visualized in figure \ref{hotcoldpopulation}, where at first (a) two bunches B1 and B3 are produced by laser pulses L1 and L3 (like in the previous scenario), but L2 comes $\Delta z \approx 4$ mm later in the lab frame and at $\xi_{\mathrm{rls,1}}=\xi_{\mathrm{rls,2}}$ the to produce B2 (b). As a result, B1 and B2 overlap in space, yet have substantially different energies due to the different acceleration times (see figure \ref{hotcoldpopulation} c) and supplemental material). Such bunch configurations are impossible with conventional accelerators. 
Slice properties of the three bunch populations of figure \ref{hotcoldpopulation} c) are plotted for the current $I$ (d), normalized emittance $\epsilon_n$ (e) and relative energy spread $\sigma_{\gamma}/\gamma$ (f). It is notable that while $I_\mathrm{B1} \approx I_\mathrm{B2}$, $I_\mathrm{B3}<I_\mathrm{B1,B2}$ although the three laser pulses are identical. This is attributed to the fact that $\xi_{\mathrm{rls,2}}$ is located much closer to the center of the blowout region, where the plasma wakefields  are significantly lower than at $\xi_{\mathrm{rls,1,2}}$. Superposition of laser and plasma fields results in reduced ionization yields and has to be taken into account when targetting a desired bunch charge.  

Simulations have been conducted using the multibunches of Fig.\ \ref{hotcoldpopulation} c)-f) to investigate the potential of the bunches to drive FEL interaction in 1-D using the Puffin FEL simulation code~\cite{campbell:093119PUFFIN}, from which the notation used below is taken. Puffin models the FEL interactions of the multibunches self-consistently within the broad bandwidth radiation fields emitted. The electron distribution evolution of the multibunches, e.g. due to energy chirps and the relative motion of the electron bunches throughout the FEL interaction, are also effectively modelled. 
An helical undulator of period $\lambda_u=1.5$ cm and r.m.s. undulator parameter $\bar{a}_u=2.5$ is used, with focusing of the multibunches provided by the `natural focusing' of the undulator. 
Relevant parameters that determine the FEL interaction for each of the three electron bunches are given in Table~\ref{FELtable} where $\left<\gamma\right>$ is the mean Lorentz factor,  $\left<\sigma_\gamma/\gamma\right>$ is the mean relative r.m.s. slice energy spread,  $\epsilon_n$ is the r.m.s. normalised emittance, $\left<\sigma_r\right>$ is the mean matched bunch radius, $\lambda_r$ is the resonant FEL wavelength, $\rho$ is the FEL parameter and $\left<\bar{Z}_R\right>=\pi F$ is the mean Rayleigh range scaled with respect to the FEL gain length $l_g=\lambda_u/4\pi\rho$, so that $F$ is the Fresnel number for a gain length ~\cite{mcnNphotonMcNeil}.

\begin {table}
\caption {FEL parameters} \label{FELtable}
\begin{center}
    \begin{tabular}{ | c  | c  | c |  c |}
    \hline
    Parameter & Bunch 1 & Bunch 2 & Bunch 3    \\ \hline
    $\left<\gamma\right>$ & 961 & 294 & 537 \\ \hline
    $\left<\sigma_\gamma/\gamma\right>$ & $1.4\times 10^{-2}$ & $4.4\times 10^{-2}$ & $1.4\times 10^{-2}$  \\ \hline
    $\left<\epsilon_n\right>$ (mm-mRad) & 0.33 & 0.37 & 0.19  \\ \hline
    $\left<\sigma_r\right>$ ($\mu$m) & 20.7 & 21.8 & 15.6  \\ \hline
    $\lambda_r$ (nm) & 59 & 630 & 189  \\ \hline
    $\rho$ & $1.05\times 10^{-2}$ & $3.35\times 10^{-2}$ & $1.73\times 10^{-2}$  \\ \hline
    $\left<\bar{Z}_R\right>$ & $0.42$  & $0.14$ & $0.12$ \\
    \hline
    \end{tabular}
\end{center}
\end{table}
   
For each of the individual bunches the normalized emittance criterion for FEL lasing of $\epsilon_n<\lambda_r\left<\gamma\right>/4\pi$ is seen to be well satisfied. When matched to the undulator focusing, these small emittances imply the matched bunch radii are relatively small. Due to these relatively dense and fine bunches the FEL coupling parameter $\rho$, at least for the lower energy bunch, approaches the limit where space-charge effects may need to be considered for more detailed modelling~\cite{Murphy1985197}. The small matched bunch radii also imply that $\left<\bar{Z}_R\right> < 1$, and radiation diffraction from the bunches will reduce the effective FEL gain length. 

An important criterion to allow efficient FEL lasing, irrespective of higher dimensional effects, is that the slice energy spread $\left<\sigma_\gamma/\gamma\right> \ll \rho$, which is not satisfied here for any of the bunches at the start of the FEL interaction. Nevertheless, as the bunches propagate through the undulator, the lowest energy bunch B2 quickly rotates and stretches in longitudinal phase space. As the bunch stretches, the slice energy spread reduces~\cite{MaierPhysRevX.2.031019} until the slice energy spread criterion becomes satisfied and FEL lasing may commence. 

\begin{figure}[htpb]
\begin{center}
  \includegraphics[width=0.99\textwidth]{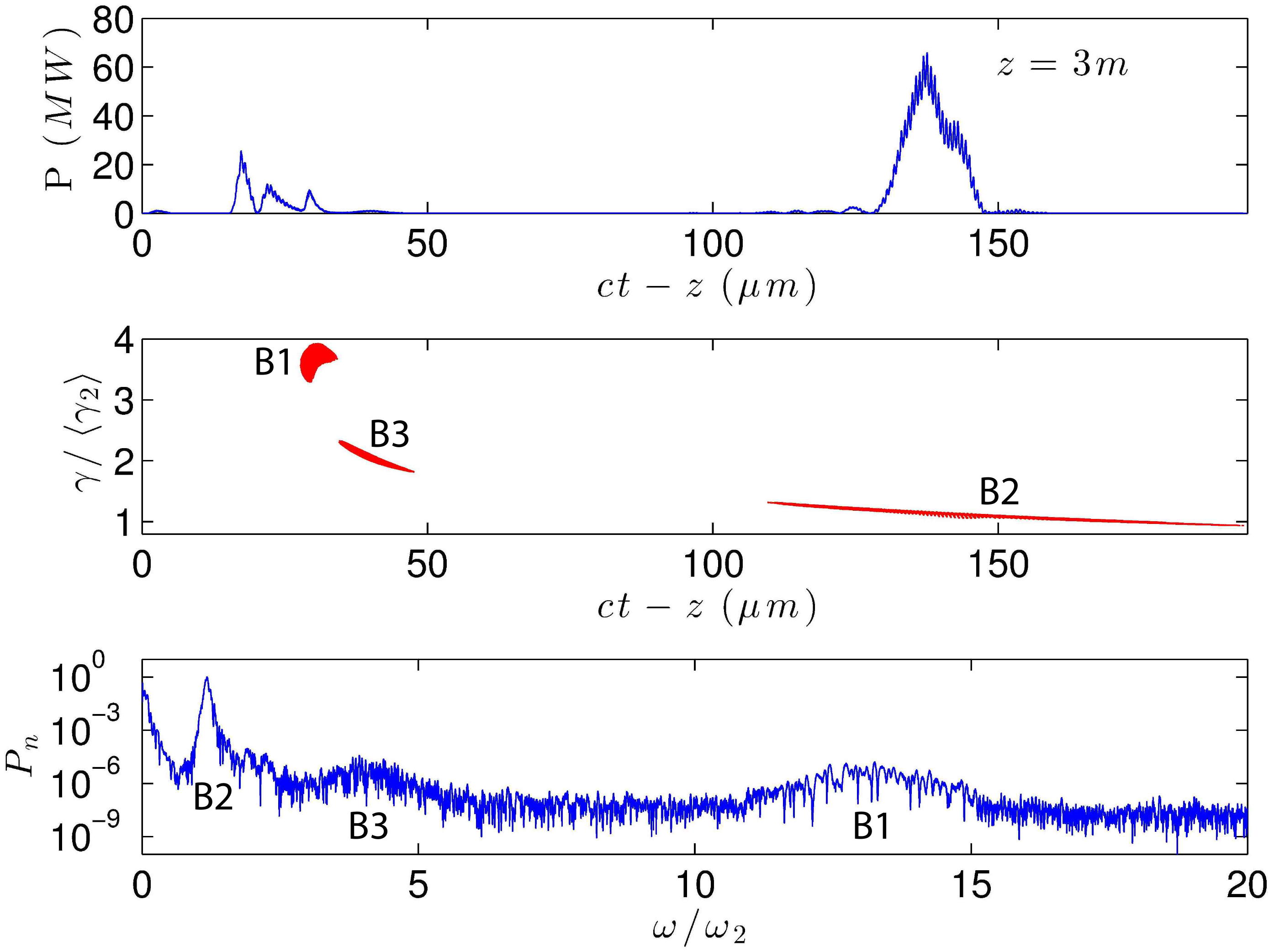}
\end{center}
\caption{Puffin 1-D simulation FEL output after 3 m undulator. Top: radiation power, middle: electron energies scaled to $\left<\gamma_2\right>$, bottom: Spectral radiation power, with respect to the peak value, $P_n=P/P_{pk}$, as a function of the frequency scaled to the mean resonant frequency  $\omega_2$ of B2.}\label{12768uns}
\end{figure}

Figure \ref{12768uns} plots the radiation output and electron phase-space distribution of the multibunches following propagation through a 3 meter undulator to investigate small signal behaviour. Comparison with the electron distributions of figure~\ref{hotcoldpopulation} shows that the lengths of the two higher energy bunches B1 and B3,  have not changed significantly from $\sim 10\,\mathrm{\mu}$m. In contrast, the low-energy B2 has been stretched to $\sim 10$ times its initial length, so reducing its slice energy spread by approximately the same factor. The energy spread criterion then becomes satisfied and FEL lasing can commence. Lasing around the maximum current of B2 is seen at $ct-z\sim 140\,\mu$m and about $\omega/\omega_2\sim 1$ in the spectral power. Smaller spectral powers are seen due to spontaneous emission from B3 and B1 about  $\omega/\omega_2\sim 4$ and $\sim 13$ respectively. Lower frequency emission $\omega/\omega_2< 0.5$ is due to coherent spontaneous emission from the multibunches. An animation of the interaction from the start of the undulator is available in the supplemental material.

The analyzed, flexible scheme shows that highly tunable electron bunches can be generated, allowing for multi-bunch trains and hitherto impossible exotic configurations such as spatially overlapping electron populations with different energies. Thanks to the strictly correlated energy chirps and high brightness of these bunches FEL-class quality can be reached in the 1D case. Further optimization of energy spread, emittance and brightness is possible and may be a unique route to achieve saturation with plasma-accelerated electron bunches. For example, the emittance can be reduced by at least an order of magnitude simply by using lower ionization threshold media such as helium. This may pave the way towards highly tunable plasma-based multi-bunch, multi-color FEL systems. 

This work was supported by DFG, by EPSRC  EP/J018171/1, STFC No. 4070022104, DOE DE-SC0009533, DE-FG02-07ER46272
DE-FG03-92ER40693, by ONR Contract No. N00014-06-1-0925 and Helmholtz VH-VI-503.
 We acknowledge the assistance of the VORPAL development team. This research used computational resources
of the National Energy Research Scientific Computing Center, which is supported by DOE DE-AC02-
05CH11231, and of JUROPA, and of HLRN. 

\bibliography{trojanmultibunch7}
\bibliographystyle{apsrev}

\end{document}